\documentclass[%
reprint,
superscriptaddress,
amsmath,amssymb,
aps,
]{revtex4-1}

\usepackage{color}
\usepackage{graphicx}
\usepackage{dcolumn}
\usepackage{bm}

\begin {document}



\title{Detrended fluctuation analysis of earthquake data
}

\author{Takumi Kataoka}
\affiliation{%
  Department of Physics, Tokyo University of Science, Noda, Chiba 278-8510, Japan
}%

\author{Tomoshige Miyaguchi}
\affiliation{%
  Department of Mathematics, Naruto University of Education, Naruto, Tokushima 772-8502, Japan
}%

\author{Takuma Akimoto}
\email{takuma@rs.tus.ac.jp}
\affiliation{%
  Department of Physics, Tokyo University of Science, Noda, Chiba 278-8510, Japan
}%



\date{\today}

\begin{abstract}
  The detrended fluctuation analysis (DFA) is extensively useful in stochastic processes to unveil the long-term correlation. 
  Here, we apply the DFA to point
    processes that mimick earthquake data. The point processes are synthesized
    by a model similar to the Epidemic-Type Aftershock Sequence model, and we
    apply the DFA to time series $N(t)$ of the point processes, where $N(t)$ is
    the cumulative number of events up to time $t$. Crossover phenomena are
  found in the DFA for these time series, and extensive numerical simulations
  suggest that the crossover phenomena are signatures of non-stationarity in the time
  series. We also find that the crossover time represents a characteristic time scale of the non-stationary process embedded in the time series. Therefore, 
  the DFA for point processes is especially useful in extracting
  information of non-stationary processes when time series are superpositions of
  stationary and non-stationary signals. Furtheremore, we apply the DFA to
  the cumulative number $N(t)$ of real earthquakes in Japan, and we
  find a crossover phenomenon similar to that found for the synthesized data.
\end{abstract}

\maketitle


\section {Introduction}
Although stationarity is one of the most important properties in stochastic
processes, non-stationary phenomena are rather ubiquitous in
nature, ranging from disordered systems \cite{Scher1975, bouchaud90,
  Bouchaud1992, Monthus1996, Brokmann2003,Metzler2014}, seismicity
\cite{Omori,Utsu1970,Ogata1988,Utsu1992} to biological systems
\cite{Weigel2011,Yamamoto2014,Manzo2015}.
Particularly in point processes, there are two typical types of non-stationary
processes. The first (non-stationarity of the first type) 
 is a process in which the probability density function (PDF) for
recurrence times depends explicitly on time. Typical examples are rainfalls that exhibit daily and seasonal
alterations. The other (non-stationarity of the second type) is a process where a characteristic time scale of the
process, such as the mean of the interval between consecutive points, diverges. Owing to divergence, the process never
reaches a steady state and thus exhibits non-stationary behaviors
\cite{bouchaud90, Metzler2014, Akimoto2020}. In this study, we focus on
non-stationary processes of the first type.

Earthquakes are an example of the non-stationary point process of
the first type. One of the most well-known statistical laws of seismicity is the
Rutenberg-Richter law \cite{GR}, which states that the magnitude distribution of
earthquakes follows an exponential distribution. This statistical law is
universal in the sense that the exponential distributions are observed in any
region on earth, any periods and any types of earthquakes such as mainshocks
and aftershocks. However, the decay constant, the so-called $b$-value, depends
on time \cite{nanjo2012decade}; thus, it is a non-stationary law.  Additionally,
the Omori law describes a non-stationary property for aftershocks \cite{Omori},
which states that the occurrence rate of aftershocks decays with the time elapsed 
from the mainshock. More precisely, the occurrence rate decays as a power law:
$\lambda (t) \propto t^{-p}$ for large $t$, where $\lambda(t)$ is the occurrence
rate at elapsed time $t$ after a mainshock, and $p$ ($>0$) is a parameter. Since
the rate of aftershocks $\lambda (t)$ explicitly depends on $t$; aftershocks are
intrinsically non-stationary, and earthquake occurrences are non-stationary
processes of the first type.


Several methods are proposed to analyze non-stationary time series. For
diffusion in heterogeneous environments, trajectories of a
diffusing particle can be tracked, and using the trajectory data, the diffusion
coefficient can be obtained from the time-averaged mean square displacement (MSD) calculated
from the trajectory.  If the process is non-stationary, the diffusion
coefficient depends explicitly on the total measurement time
\cite{He2008,Weigel2011,Metzler2014,Yamamoto2014,Miyaguchi2011,Miyaguchi2015,AkimotoYamamoto2016a}.
Thus, plotting the diffusion coefficient as a function of the measurement time
provides us information on how the process ages. In another method of
non-stationary data analysis of the second type, the inter-occurrence times are
 utilized frequently \cite{Wong2004,Weigel2011,kuno2000nonexponential}.  For example, the inter-occurrence-time PDFs have also been 
 extensively used for earthquake researches \cite{Corral2004, abe2005scale, Saichev2006, hasumi2009weibull,
  tanaka2017detailed}.


Although the time-averaged MSD and inter-occurrence-time PDF are useful
analysis methods for non-stationary time series of the second type, these methods are 
not effective in case of the first type. In non-stationary processes of
the first type, a long-measurement-time limit of a time average may have a
definite value, as the process does not age monotonically.  Thus, the
diffusion coefficient of time-averaged MSD does not depend on the total
measurement time. Moreover, the inter-occurrence-time PDF analysis is based on
the fact that the inter-occurrence times are independent and identically
distributed (IID). However, this assumption is not valid for  nonstationary
processes of the first type.  Therefore, it is important to develop a method to
extract information of nonstationary features from the time series.

In this study, we utilize the detrended fluctuation analysis (DFA), to overcome the above-mentioned difficulties of non-stationary
time-series analysis of earthquakes \cite{Peng1994}.
The difference between the DFA and the time-averaged MSD is  that the local trends
are subtracted in the DFA. The DFA of recurrence
times of earthquakes in stationary regimes was studied to unveil the long-term
correlation \cite{lennartz2008long}. We note that the DFA can be
utilized irrespective of the inter-occurrence times being IID. Here, we
apply the DFA to both data synthesized by an earthquake model and data of real
earthquakes in non-stationary regimes. In particular, we perform the DFA to a
cumulative number $N(t)$ of earthquakes occurred up to time $t$.  As an
earthquake model, we employ a simplified version of the Epidemic-Type aftershock
sequence model \cite{Ogata1988}.

We find crossover phenomena in DFA for both synthesized and real earthquake
data. It is shown that the crossover time represents a characteristic time scale
of non-stationary process.  Moreover, we present analytical predictions of
long time behaviors in DFA for point processes. Until now, relations between the
DFA and the long-term correlation in time series have been analytically obtained
for fractional Brownian motion (fBm) \cite{taqqu95}, but it is important to obtain 
an analytical expression for point processes as well, as these two are totally different
stochastic processes \cite{magdziarz09}. The DFA has
been widely used in data analysis, and thus the analytical results for
point processes are also useful.


This article is organized as follows. In Sec.~\ref{s.earthq-model}, the
earthquake model, which is a superposition of stationary and non-stationary
point processes, is proposed, and in Sec.~\ref{s:dfa}, the DFA method is
briefly reviewed. Sections \ref{s:dfa-for-model} and \ref{s:dfa-for-realdata}
present results of DFA for synthesized data and real earthquake data,
respectively. Finally, Sec.~\ref{s.conclusion} presents a summary and
discussion.

\section {Earthquake model}\label{s.earthq-model}

Here, we propose a point process describing occurrences of earthquakes over a vast
area, such as the entire extent of Japan. In our model, three types of earthquakes
are considered: mainshocks, aftershocks, and stationary earthquakes independent
of mainshocks and aftershocks (background earthquakes).

\begin{figure}
\includegraphics[width=.9\linewidth, angle=0]{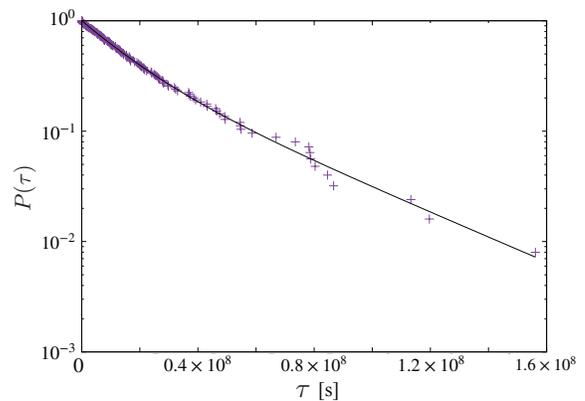}
\caption{Survival probability $P(\tau)$ of inter-occurrence times
  of mainshocks (semi-log plot).  There have been 135 mainshocks from January 1, 1923
  to April 30, 2017 in Japan. Crosses are a result of real earthquake data.
  The solid line represents a superposition of the exponential
    distributions, i.e.,
    $P(\tau) = A \exp(-\lambda_1 \tau) + (1-A) \exp(-\lambda_2 \tau)$, with
    $A=0.59$, $\lambda_1^{-1} \cong 1.48 \times 10^7$ [s] and
    $\lambda_2^{-1} \cong 3.87 \times 10^7$ [s].  }
\label{sp-main}
\end{figure}

First, we assume that mainshocks occur independently, and thus they are
described by a Poisson process. This assumption is quite reasonable because the
superposition of large numbers of mutually independent renewal processes becomes
a Poisson process in general \cite{Cox}. In fact, mainshock occurrences have
been considered a Poisson process \cite{gardner1974sequence,
  kagan1991long}.
We have partially confirmed this assumption for a region around Japan by
analyzing the Japan Meteorological Agency (JMA) catalog \footnote{Japan
  Meteorological Agency Earthquake Catalog,
  http://evrrss.eri.u-tokyo.ac.jp/tseis/jma1/index.html}, where we define the mainshocks as
earthquakes with magnitudes greater than 7. Figure~\ref{sp-main} shows that
survival probability $P(\tau)$ of inter-occurrence times $\tau$ between
successive mainshocks follows a superposition of exponential distributions. In
the data, the mean inter-occurrence time is around $2.20\times 10^7$ [s]
$\cong 254$ [d]. The exponential distribution with mean $2.20\times 10^7$ [s]
also well describes the inter-occurrence distribution.  Therefore, the mainshock
rate $\lambda_m$ for the entire extent of Japan is approximately given by
$\lambda_m \cong 1/ 2.2\times 10^{-7}$ [1/s].

Second, we assume that aftershocks are triggered by a mainshock. In particular,
we assume that the Omori law \cite{Omori}, which states that the rate of occurrence
of aftershocks after a mainshock follows a power-law decay:
\begin{equation}
  \label{e.lambda_a(t)}
\lambda_a (t) = \frac{K}{(t+c)^p},
\end{equation}
where $t$ is the elapsed time after a mainshock, and $K$ is the degree of the aftershock activity, $c$ is a parameter characterizing the relaxation time of the activity, and $p$
is the power-law exponent. I
In particular, it has been shown that the parameter $p$ clearly
depends on the magnitude of the mainshock \cite{ouillon2005magnitude}.

\begin{figure}
\includegraphics[width=1.\linewidth, angle=0]{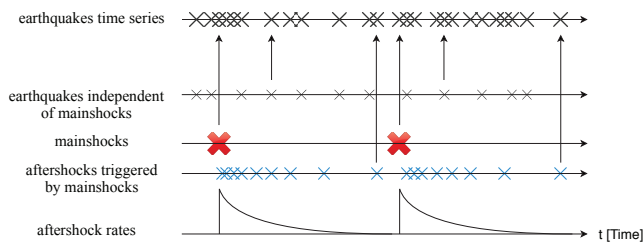}
\caption{Schematic view of earthquake model. Earthquake time series are obtained
  by superposing mainshocks, aftershocks, and background earthquakes. We assume
  that aftershock rate $\lambda_a(t)$ is the same for each mainshock.  }
\label{model}
\end{figure}

Third, we assume that background earthquakes occur independently of mainshocks
and aftershocks. This assumption differs from the Epidemic-Type Aftershock
Sequence model \cite{Ogata1988}, where every earthquake is triggered by
mainshocks or aftershocks, which are usually triggered by a mainshock.  Furthermore, we
  assume that the rate of the background earthquakes depends on the magnitudes
  of the mainshocks; that is, when the magnitude of the mainshock is large, the rate of
  the background earthquakes is also large.
It is difficult to determine whether an earthquake is an aftershock or a
background earthquake.  As we consider earthquakes over a vast region, we
assume that almost all earthquakes are independent of mainshocks, and thus are 
background earthquakes. Under this assumption, we can determine some of the
model parameters.
Figure~\ref{model} shows a schematic view of our model. 

A Poisson process with rate $\lambda$ can be generated by creating inter-event
times following the exponential distribution with mean $1/\lambda$ in numerical
simulations.  In particular, the mainshocks and background earthquakes are
generated by Poisson processes with rates $\lambda_s$ and $\lambda_m$,
respectively. As aftershocks are described by the Omori's law, we generate
them by using a non-stationary Poisson process (see Appendix~A for the details).
Unlike a non-Markov model of aftershocks \cite{akimoto2005large}, it is a Markov
model with the exception that it is non-stationary.

\section {Detrended fluctuation analysis}\label{s:dfa}

The DFA was invented to analyze data that have local trends and non-stationary
features \cite{Peng1994}. This method has been used to unravel long-range
correlations in stationary as well as non-stationary time series such as
heartbeat rates, weather variations, recurrence times of earthquakes, and
conformation fluctuations of proteins
\cite{Peng1993,Bunde1998,havlin1999scaling,lennartz2008long,harada09,Yamamoto2014b}.

The basic idea is to quantify fluctuations around local trends as
\begin{equation}
F^2(n) \equiv \frac{1}{mn} \sum_{j=0}^{m-1} \sum_{i=jn+1}^{(j+1)n} (y_i - \tilde{y}_i^j)^2,
\label{def-dfa}
\end{equation}
where $y_i$ is a time series that are considered, and $\tilde{y}_i^j$
represents a local trend in the time interval $[jn+1, (j+1)n]$. Thus, $n$ is the
length of these time intervals. The local trend $\tilde{y}_i^j$ in the interval
$i \in [jn+1, (j+1)n]$ is given by a linear function obtained by the
least-square fit to data $y_i$ in the same interval.

Even when there are local trends in data, the function $F(n)$ characterizes a
long-term correlation in the data. In particular, $F(n)$ increases as
$F(n) \propto n^{1/2}$ when there is no correlation in increments $\Delta y_i$
of $y_i$, i.e., $\Delta y_i \equiv y_{i} - y_{i-1}$. However, it
increases as $F(n) \propto n^{\alpha}$ with $\alpha \ne 1/2$ when the increment
has a strong correlation, implying a power-law decay of the correlation function.  In
particular, $\alpha <1/2$ implies that there is an anti-correlation in
increments $\Delta y_i$ and $\alpha >1/2$ implies a positive correlation of
increments $\Delta y_i$

Here, we apply the DFA to time series $y_i$ generated by a point process. More
precisely, $y_i$ is a monotonically increasing sequence defined by $y_i=N(i)$,
where $N(i)$ is the cumulative number of earthquakes up to time $i$. The
variable $i$ is an integer in the original DFA, whereas $i$ in $N(i)$ represents the
continuous time; thus, it is a real number. Hence, in what follows, we use $t$
as the argument and use the notation $N(t)$. Note, however, that the definition of
the function $F(n)$ in Eq.~(\ref{def-dfa}) remains unchanged even for point
processes because we only use discrete data points of $N(t)$.  In a previous
study \cite{paradisi2012scaling}, the DFA for a sequence generated by a point
process such as $N(t)$ was studied, where inter-occurrence times are IID random
variables. Such a process is called a renewal process.  However, the
inter-occurrence times may not  be IID and the time series are non-stationary of the
first type in earthquakes. Here, we investigate the DFA for point processes for such
non-stationary time series.

\section {Detrended fluctuations analysis on synthesized data}\label{s:dfa-for-model}

In this section, the DFA is applied to three types of data synthesized by the
earthquake model: (1) Background earthquakes (Poisson processes), (2) One
mainshock and its aftershocks without background earthquakes, and (3)
Poissonian mainshocks with aftershocks and background earthquakes. Numerical
simulations are carried out for these models and compared with theoretical
predictions for small and large $n$. Derivations of these predictions are
presented in Appendices \ref{s.theory-dfa-background} and
\ref{s.theory-dfa-aftershocks}.

\subsection {Background earthquakes (Poisson process)}

\begin{figure}
\includegraphics[width=.9\linewidth, angle=0]{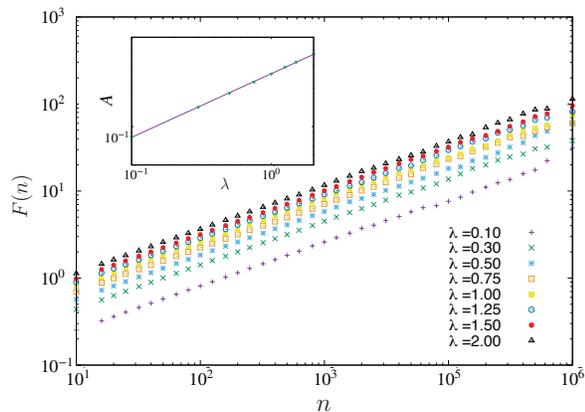}
\caption{Detrended fluctuation analysis of $N(t)$ in Poisson processes for
  different $\lambda$, where the total length of the time series is fixed at $10^7$.
  Inset: constant $A$ as a function of $\lambda$. The solid line represents
  $A=\sqrt{\lambda}/\sqrt{15}$}.
\label{dfa-poisson}
\end{figure}

First, we apply the DFA to Poisson processes, for which inter-occurrence times
follow an exponential distribution with rate $\lambda$.
Figure~\ref{dfa-poisson} shows that $F(n)$ increases as $F(n)=An^{1/2}$ for any
$n>0$ and a constant $A$ depends on rate $\lambda$ of the Poisson process.  A
theory of the DFA for a Poisson process implies
\begin{equation}
  \label{e.dfa_poisson_coef}
  A \cong
  \sqrt{\frac {\lambda}{15}}
\end{equation}
(a proof is given in Appendix \ref{s.theory-dfa-background}), which is confirmed
using numerical simulations (inset of Fig.~\ref{dfa-poisson}).  As a
Poisson process is a memory-less process, the scaling of $F(n) \propto n^{1/2}$
is quite reasonable. However, we numerically find that scaling
$n^{1/2}$ is no longer valid for renewal processes where the PDF of
inter-occurrence times follows a power-law distribution with a divergent mean.
Therefore, scaling $n^{1/2}$ represents the signature of a stationary Poisson
process.  In a biased continuous-time random walk, the variance of the
displacement, which is a quantity similar to the DFA, shows an anomalous scaling
\cite{Akimoto2018b, Hou2018}.

\begin{figure*}
\includegraphics[width=.95\linewidth, angle=0]{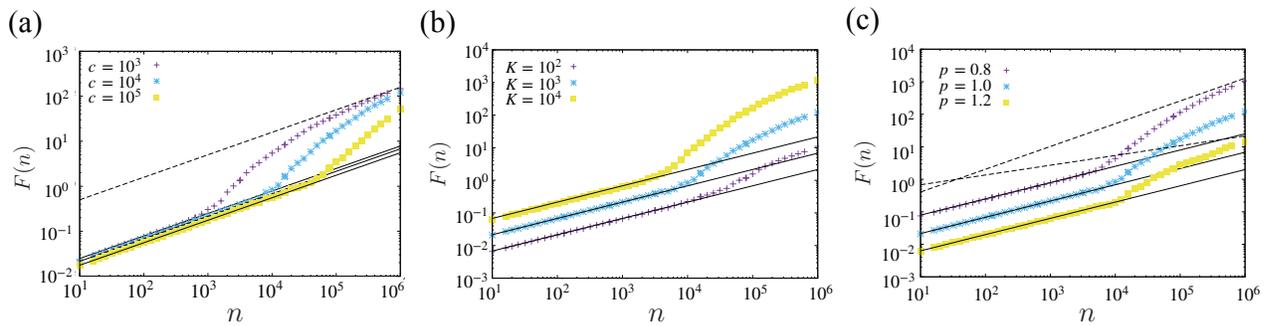}
\caption{Detrended fluctuation analysis (DFA) of $N(t)$ for earthquake model,
  where the length of time series is fixed as $10^7$.  (a) DFAs for different values of 
  $c$. The other parameters are fixed at $p=1$ and $K=10^3$. All the solid lines
  represent $F(n)=Bn^{1/2}$ using Eq.~(\ref{B}). The dashed line
  represents Eq.~(\ref{e.F(n).p=1}).  (b) DFAs for different $K$ values. The other
  parameters are fixed at $p=1$ and $c=10^4$.  (c) DFAs for different values of $p$. The other
  parameters are fixed at $c=10^4$ and $K=10^3$.  All the solid lines represent
  $F(n)=Bn^{1/2}$ using Eq.~(\ref{B}). The dashed lines
  represent Eq.~(\ref{e.F(n).p<3/2}).  }
\label{dfa-models}
\end{figure*}

\subsection {One mainshock and its aftershocks without background earthquakes}

Second, synthesized data $N(t)$ are generated using the earthquake model. To obtain
a deeper understanding of the features of the DFA, we consider a simple situation
where a mainshock occurs only once at $t=0$, and there are no background
earthquakes, i.e., the time series comprises one mainshock and its aftershocks.

As shown in Fig.~\ref{dfa-models}, all the results of the DFA show a crossover
from $n^{1/2}$ to $n^\alpha$ scaling.  For small-$n$ behavior, $F(n)$ shows
$F(n)=B n^{1/2}$. By an adiabatic approximation, we approximately obtain $B$:
\begin{equation}
B \cong \frac{\sqrt{\overline{\lambda_a}}}{\sqrt{15}},
\label{B}
\end{equation}
where 
\begin{equation}
\overline{\lambda_a} = \frac{1}{T} \int_0^T \lambda_a(t)dt
\end{equation}
and $T$ is the total length of the time series. For large-$n$ behavior, $F(n)$ also
shows
\begin{equation}
  \label{e.F(n).p=1}
F(n) \sim \frac{K}{2\sqrt{T}} n^{1/2}
\end{equation}
when $p=1$ (see Appendix.~\ref{s.theory-dfa-aftershocks}).
Equation (\ref{e.F(n).p=1}) is a special case of a general result for $p<3/2$
\begin{equation}
  \label{e.F(n).p<3/2}
  F(n) \sim \frac{Kp}{\sqrt{T(3-2p)} (3-p)(2-p)}  n^{3/2 -p},
\end{equation}
which is valid for $n\to \infty$ (see
Appendix.~\ref{s.theory-dfa-aftershocks}).  Thus, the power-law
  exponent in the DFA is determined by $p$. In other words, the parameter $p$
  can be obtained from the asymptotic behavior of the DFA for $N(t)$. This is
  one of the most important analytical results of our study.

  Figure~\ref{dfa-models} summarizes the results of the DFAs of $N(t)$ for
  different parameters. Figure~\ref{dfa-models}(a) shows that crossover time
  $n_c$ in $F(n)$ increases with increasing parameter $c$ and short-$n$
  behaviors are almost the same. For large-$n$ behavior, $F(n)$ converges to a
  $n^{1/2}$ scaling, which does not depend on $c$.  As shown in
  Figs.~\ref{dfa-models}(b) and (c), crossover time $n_c$ also depends  on $K$
  and $p$, but the dependencies are relatively weak compared with the $c$
  dependency. Importantly, crossover time $n_c$ is thus almost proportional to
  $c$. Therefore, the parameter $c$ can be estimated from the crossover time
  $n_c$. This is significantly important when time series are a superposition of
  non-stationary and stationary signals, because information of the
  non-stationary part can be obtained without distinguishing the stationary and non-stationary
  time series. In Fig.~\ref{dfa-models}, it is clearly shown that the asymptotic
  behaviors of $F(n)$ exhibit different power-law scaling with exponent
  $3/2-p$. Therefore, the parameter $p$ can be obtained from the asymptotic
  behavior of the DFA for $N(t)$ if background earthquakes are removed from the
  time series.


\subsection {Poissonian mainshocks with aftershocks and background earthquakes}

For synthesized data $N(t)$ generated by the earthquake model with
  several mainshocks, we find a crossover phenomenon such that $F(n)$ exhibits a $n^{1/2}$ to $n^{\alpha}$ scaling.  The superposition of two Poisson processes with
rates $\lambda_1$ and $\lambda_2$ is equivalent to a Poisson process with rate
$\lambda_1+ \lambda_2$. Therefore, the DFA for the superposition of the two Poisson
processes with rates $\lambda_1$ and $\lambda_2$ becomes
$F(n)=\sqrt{(\lambda_1+ \lambda_2)n} /\sqrt{15}$.  The synthesized time series
are composed of background earthquakes and aftershocks triggered by a
mainshock. Since background earthquakes and mainshocks are described by Poisson
processes with rates $\lambda_b$ and $\lambda_m$, the above estimation can be
utilized.  For small-$n$ behavior, $F(n)$ shows that $F(n) \simeq C n^{1/2}$ and $C$ can be
approximately obtained as
\begin{equation}
C \cong \frac{\sqrt{\lambda_b + \lambda_m \times N_a}}{\sqrt{15}},
\end{equation}
where 
\begin{equation}
N_a = \int_0^{1/\lambda_m} \lambda_a(t)dt.
\end{equation}
The small-$n$ behavior of the DFA for $N(t)$ is determined by $\lambda_b$, $\lambda_m$, and $N_a$. 

\begin{figure*}
\includegraphics[width=.9\linewidth, angle=0]{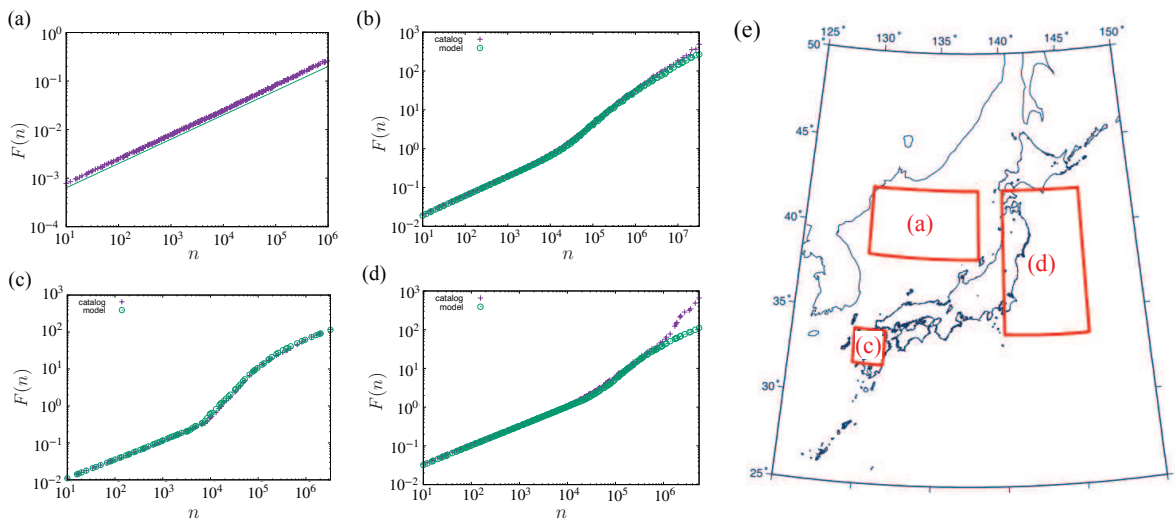}
\caption{Detrended fluctuation analysis (DFA) of $N(t)$ for earthquake
  data. Crosses are the results for the catalog data (real earthquakes), and circles
  are those for data synthesized with the earthquake models.  
  (a) DFA for time series $N(t)$ with no mainshock at someplace. $N(t)$ is obtained with
  earthquakes occurred inside the area enclosed within 38${}^\circ$-42${}^\circ$
  N latitude and 130${}^\circ$-138${}^\circ$ E longitude [see subplot (e)]. The
  period of time series $N(t)$ is restricted from January 1, 2009, to December
  31, 2010.  
  (b) DFA for time series $N(t)$ with several mainshocks. $N(t)$ is obtained
  with earthquakes occurred inside the area enclosed within
  25${}^\circ$-50${}^\circ$ N latitude and
    125${}^\circ$-150${}^\circ$ E longitude. The parameters of the earthquake model
    are $\lambda_s=5.0 \times 10^{-4}$, $c=10^4$, $K=10^3$, and $p=1.15$.  The
  period of time series $N(t)$ is restricted from January 1, 2001, to December
  31, 2010. 
  (c) DFA for Kumamoto earthquakes. $N(t)$ is obtained with earthquakes occurred
  inside the area enclosed within 32${}^\circ$-34${}^\circ$ N latitude and
  129.5${}^\circ$-131.5${}^\circ$ E longitude [see subplot (e)]. The period of
  time series $N(t)$ is restricted from April 16, 2016, to April 30, 2017.
  The parameters of the earthquake model are
    $\lambda_s=1.5 \times 10^{-5}$, $c=10^4$, $K=5.0 \times 10^3$, and
    $p=1.15$.  
  (d) DFA for Tohoku earthquakes. $N(t)$ is obtained with earthquakes occurred
  inside the area enclosed within 34${}^\circ$-42${}^\circ$ N latitude and
  140${}^\circ$-146${}^\circ$ E longitude [see subplot (e)]. The period of time
  series $N(t)$ is restricted from March 11, 2011, to December 6, 2012.
  The parameters of the earthquake model are
    $\lambda_s=1.5 \times 10^{-3}$, $c=3.0 \times 10^4$, $K=10^4$, and $p=1.2$.
  (e) Three regions  used for the analysis of the subplots (a)(c)(d). The entire region is used for the analysis of the subplot
  (b).  }
\label{dfa-catalog}
\end{figure*}
 
\section {Detrended fluctuation analysis on earthquake data catalog}\label{s:dfa-for-realdata}

Here, we apply the DFA to the cumulative number $N(t)$ of real earthquakes
included in the JMA catalog; this catalog contains data of earthquakes with
magnitude $M\geq 2$ and the ones that occurred in the area of 25${}^\circ$-50${}^\circ$ N
latitude and 125${}^\circ$-150${}^\circ$ E longitude.  Figure~\ref{dfa-catalog}
shows DFAs for $N(t)$ for several different periods and areas in Japan.  We find
a crossover phenomenon in $F(n)$; i.e., $F(n)$ increases as
$F(n)\propto n^{1/2}$ for small $n$ and $F(n)$ shows another scaling for large
$n$ if the data are affected by at least one mainshock. More precisely, $F(n)$
increases as $F(n) \propto n^\alpha$ for $n>n_c$, where $\alpha \cong 1.0$ and
$n_c \cong 1.6 \times 10^4$ [s] ($=4.5$ h).  This crossover is observed for
all earthquake time series if they contain mainshocks, i.e., earthquakes with
magnitudes greater than 7.  Moreover, small-$n$ behaviors of $F(n)$ are almost
equivalent for all data, while the crossover times $n_c$ are slightly
different. Furthermore, we observe that synthesized data $N(t)$ generated by the earthquake
model, where parameters are set to be
$\lambda_s^{-1}=2.0 \times 10^3$ [s], $c=10^4$ [s], $K=10^3$, and
  $p=1.15$, show similar crossover phenomena in the DFA [see
  Fig.~\ref{dfa-catalog}(b)]. In this synthesized data, instead of generating
  mainshocks according to Poisson statistics, mainshocks are assumed to occur at
  the same times $t_i\,(i=1,2,\dots)$ at which real earthquakes with magnitudes
  greater than $7$ occurred.


Furthermore, we applied the DFA to $N(t)$ for specific earthquakes such as the Tohoku
and Kumamoto earthquakes.  In the Tohoku earthquakes, a mainshock occurred on March 11, 2011,
  with  a magnitude of $M=9.0$ and we analyzed earthquakes after the mainshock whose area overlaps the area around the epicenter 
  [see  Fig.~\ref{dfa-catalog}(e)].  In the Kumamoto
  earthquakes, a mainshock occurred on April 16, 2016, with a magnitude of $M=7.3$, and
  we analyzes earthquakes after the mainshock whose area overlaps the area around the epicenter [see Fig.~\ref{dfa-catalog}(e)]. In the DFAs for these
  real earthquakes [Fig.~\ref{dfa-catalog}(c)(d)], we find crossover phenomena
  similar to that found for the synthesized data. Moreover, we successfully
generate time series $N(t)$ with our earthquake model that reproduce the DFAs of
the two earthquake time series, i.e., the Tohoku and Kumamoto earthquakes
[circles in Fig.~\ref{dfa-catalog}(c)(d)]. In the DFA of the Tohoku
  earthquakes at large $n$, however, there is a slight difference between the
  results of the catalog data and the earthquake model. While we assume that a
  mainshock occurs only at $t=0$, there are a few mainshocks (earthquakes with
  magnitudes greater than 7) occurring after $t=0$ in the catalog data.  We believe that
  such large aftershocks significantly affect subsequent aftershocks. Therefore,
  our model cannot completely reproduce the DFA of Tohoku earthquakes.

\section {Conclusion}\label{s.conclusion}

We found that crossover phenomena in the DFA of $N(t)$, i.e., the number of
earthquakes up to time $t$, are universally observed in earthquake
data. Extensive numerical simulations of the earthquake model show that the
crossover phenomena originate from the non-stationarity of the aftershock
sequences. In particular, crossover time $n_c$ in the DFA characterizes
parameter $c$, which represents the relaxation time of aftershocks in
  the Omori's law. Although we do not determine if an earthquake is an
  aftershock or not, we can successfully obtain information regarding the 
  aftershocks. Therefore, our analysis is significantly important when the time
  series is a superposition of the two types of time series that cannot be distinguished and when one of the two types is non-stationary and the other is stationary. Moreover, we present theories of the DFA for stationary and
  non-stationary point processes, which are necessary for performing a thorough analysis.

\section* {Acknowledgement}
T.A. was supported by JSPS Grant-in-Aid for Scientific Research (No.~C JP18K03468).

\appendix
\section {Numerical simulations of non-stationary Poisson processes}

In numerical simulations of the earthquake model in
  Sec.~\ref{s.earthq-model},  aftershock sequences, which
  follow a non-stationary Poisson process, must be generated. Aftershocks triggered by a mainshock
  are generated as follows:  Here, we assume that a mainshock occurs at $t=-t_0$
  ($t_0>0$). First, time $t_0$ can be determined as follows:  As mainshocks are
  described by a Poisson process with rate $\lambda_m$, $t_0$ can be obtained by
  generating a random variable following the exponential distribution with rate
  $\lambda_m$. This is because the random variable $t_0$, which is called the
  backward recurrence time in renewal theory \cite{Cox}, follows the same exponential
  distribution as the inter-occurrence-time distribution of mainshocks as a result of the memory-less property of Poisson processes \cite{Cox}.  In computer programs, $t_0$ is obtained by $t_0=- \lambda_m^{-1} \ln X$, where
  $X$ is a random variable uniformly distributed on $[0,1]$.

  The probability that an aftershock triggered by the mainshock at $t=-t_0$
  occurs at $t=0$ is given by $\lambda_a(t_0) \Delta t$, where $\Delta t$ is a
  time step. In the JMA earthquake catalog, time step $\Delta t$ is $\Delta t=1$
  [s]. In computer programs,  a random variable $X$ is uniformly distributed on
  $[0,1]$. Then, we generate an aftershock at $t=0$  if $X<\lambda_a(t_0) \Delta t$, and no aftershock occurs at $t=0$ otherwise. This procedure is repeated for
  $t=\Delta t, 2\Delta t, \dots$ with rates
  $\lambda_a(t_0+\Delta t), \lambda_a(t_0+2\Delta t), \dots$.  In particular,
  the probability that an aftershock triggered by the mainshock occurs at
  $t=n\Delta t$ is given by $\lambda_a(t_0 + n \Delta t) \Delta t$.

For simplicity, we generate aftershocks triggered by the mainshock at $t=-t_0$
until the next mainshock occurs at $t=t_1$.  This is because the rate of
aftershocks triggered by the next mainshock $\lambda_a(t-t_1)$ is much greater
than that triggered by the previous mainshock $\lambda_a(t+t_0)$ where $t>t_1$.
Similar procedures were employed for the subsequent mainshocks at
$t=t_2, t_3, \dots$ and their aftershocks.

\section {Dimensionless form of occurrence rate}

Here, we transform the time-dependent occurrence rate $\lambda(t)$
  [Eq.~(\ref{e.lambda_a(t)})] in a dimensionless form.  By the following
  transformation,
\begin{equation}
  t \to t/c \equiv \tilde{t},
  ~\lambda_a(t) \to c \lambda_a (t) \equiv \tilde{\lambda}_a(\tilde{t}),
  ~K\to c^{1-p}K \equiv \tilde{K},  
\end{equation} 
we have a non-dimensional occurrence rate
\begin{equation}
  \tilde{\lambda}_a (\tilde{t}) = \frac{\tilde{K}}{(\tilde{t} + 1)^p}. 
\end{equation}
In addition, if the measurement time $T$ is transformed as
$T\to T/c \equiv \tilde{T}$,  the remaining parameters are $\tilde{K}, \tilde{T}$
and $p$.

\section {Theory of DFA for point process}\label{s.theory-dfa-background}

In this study, our objective is the extraction of nonstationary information from
point processes using DFA.  Here, we provide a theoretical argument regarding the
DFA for stationary point processes with constant rate $\lambda$.

\subsection {Inter-occurrence-time distribution}

We assume that inter-occurrence-time distribution of successive renewals follows
a distribution $\psi(\tau)$ with finite mean and variance $\langle \tau \rangle$
and $\langle \tau^2 \rangle - \langle\tau\rangle^2$. For this
inter-occurrence-time distribution, the mean and variance of $N(i)$ are given by
\cite{God2001,Cox}
\begin{align}
  \label{e.<N(i)>}
  \left\langle N(i) \right\rangle
  &\sim \frac{i}{\left\langle \tau \right\rangle},
  \\[0.1cm]
  \label{e.<N(i)^2>c}
  \left\langle N^2(i) \right\rangle - \left\langle N(i) \right\rangle^2
  &\sim
  \frac{\left\langle \tau^2 \right\rangle - \left\langle \tau \right\rangle^2}
  {\left\langle \tau \right\rangle^3} i,  
\end{align}  
In particular, mean interval $\langle\tau\rangle$ is related with the rate
$\lambda$ as $\langle\tau\rangle = 1/\lambda$; thus, Eq.~(\ref{e.<N(i)>}) is
rewritten as $\langle N(i) \rangle \sim \lambda i$. For Eq.~(\ref{e.<N(i)^2>c}),
we also use a notation
$\langle N^2(i) \rangle - \langle N(i) \rangle^2 = \sigma^2i$, where $\sigma^2$
is defined by
$\sigma^2=(\langle\tau^2\rangle - \langle\tau\rangle^2)/\langle \tau\rangle^3 $.

If $\psi(\tau)$ is given by the exponential distribution
\begin{equation}
 \psi(\tau) = \frac {1}{\left\langle \tau \right\rangle} e^{-\tau/\langle\tau\rangle},
\end{equation}
the point process is referred to as the Poisson process. For the Poisson process
Eqs.~(\ref{e.<N(i)>}) and (\ref{e.<N(i)^2>c}) is given by
\begin{align}
  \label{e.<N(i)>.poisson}
  \left\langle N(i) \right\rangle
  &= \frac{i}{\left\langle \tau \right\rangle},\quad
  \\[0.1cm]
  \label{e.<N(i)^2>c.poisson}
  \left\langle N^2(i) \right\rangle - \left\langle N(i) \right\rangle^2
  &= \frac{i}{\left\langle \tau \right\rangle}.
\end{align}  
Note that equalities hold for the Poisson process. From the definition of
$\sigma^2$, it  follows that $\sigma^2 = \lambda$ holds for the Poisson
process.

\subsection {Theory of DFA for point process}

Based on the stationarity of the point process, Eq.~(\ref{def-dfa}) can be represented as
\begin{equation}
F^2(n) = \left\langle \frac{1}{n} \sum_{i=1}^n (y_i - \tilde{y}_i^0)^2 \right\rangle,
\end{equation}
where $y_i=N(i)$ is obtained using the point process described above.  We rewrite the
DFA as
\begin{eqnarray}
  F^2(n) &=& \frac{1}{n} \sum_{i=1}^n
             \left\langle [y_i - \lambda i - (\tilde{y}_i^0 - \lambda i) ]^2 \right\rangle\\
         &=& \frac{1}{n} \sum_{i=1}^n [\tilde{y}_i - (ai +b)]^2,
\end{eqnarray}
where $\tilde{y}$ is defined as $\tilde{y}_i = y_i - \lambda i$. Moreover, $a$
and $b$ are the coefficients of the linear fitting of $\tilde{y}_i$ by the
least-square method, i.e., $\tilde{y}_i^0 - \lambda i = a i + b$. By expanding
the summand, we obtain
\begin{eqnarray}
  F^2(n) &=& \frac{1}{n} \sum_{i=1}^n
             (\left\langle \tilde{y}_i^2 \right\rangle
             -2i \langle a \tilde{y}_i \rangle
             -2 \langle b \tilde{y}_i \rangle
             \nonumber\\
  \label{e.F(n)^2}
         && + i^2\langle a^2 \rangle + 2i \langle ab \rangle + \langle b^2 \rangle).
\end{eqnarray}

The parameters $a$ and $b$ are given by
\begin{equation}
  \label{e.a_and_b}
  a= \frac{S_{xy}}{S_x^2}- \lambda, \qquad
  b=\frac{1}{n} -\frac{n+1}{2} \frac{S_{xy}}{S_x^2}, 
\end{equation}
where $S_{xy}$ and $S_x^2$ are a covariance and a variance given by
\begin{align}
  \label{e.Sxy}
  S_{xy}  &=
  \frac{1}{n} \sum_{j=1}^n j y_j -\frac{1}{n^2}\sum_{j=1}^n j\sum_{k=1}^n  y_k,
  \\
  \label{e.Sx^2}
  S_{x}^2 &=
  \frac{1}{n} \sum_{j=1}^n j^2 -\frac{1}{n^2} \left(\sum_{j=1}^n j \right)^2 .
\end{align}
It follows that $a$ and $b$ are written as
\begin{align}
  \label{e.a}
  a&= \frac{1}{nS_x^2 } \sum_{j=1}^n \left( j - \frac{n+1}{2}\right) \tilde{y}_j
  \\[0.0cm]
  \label{e.b}
  b&= \frac{1}{n} \sum_{j=1}^n \tilde{y}_j - \frac{n+1}{2}a.
\end{align}

In Poisson processes, we have
$\langle N(i) N(i+i') \rangle = \langle N^2(i) \rangle + \langle N(i) [N(i') -
N(i) ]\rangle = \langle N^2(i) \rangle + \langle N(i)\rangle \langle N(i') -
N(i) \rangle$, because a Poisson process is a memory-less process, i.e., $N(i)$
and $ N(i') - N(i) $ are independent.  It follows that
$\langle \tilde{y}_i \rangle =0$, $\langle \tilde{y}_i^2 \rangle =\sigma^2 i$,
and $\langle \tilde{y}_i \tilde{y}_j \rangle = \sigma^2 \min (i,j)$. For general
point processes, $N(i)$ and $ N(i') - N(i) $ are not independent. It becomes
\begin{equation}
  \label{e.<yiyj>}
  \langle \tilde{y}_i \tilde{y}_j \rangle \approx \frac{i+j}{2}\sigma^2 - \frac{|i-j|}{2}\lambda.
\end{equation}

 \begin{widetext}
By using Eqs.~(\ref{e.a}) and (\ref{e.<yiyj>}), we obtain
\begin{equation}
  \label{e.<y_ia>}
  \sum_{i=1}^n \left(i-\frac{n+1}{2}\right)\langle \tilde{y}_i a \rangle
  \approx \frac{-\lambda}{nS_x^2 } \sum_{i,j=1}^n \left( i - \frac{n+1}{2}\right)\left( j - \frac{n+1}{2}\right) \frac{|i-j|}{2},
\end{equation}
and
\begin{equation}
  \label{e.<a^2>}
\langle  a^2 \rangle = \frac{-\lambda}{(nS_x^2)^2 } \sum_{i,j=1}^n \left( i - \frac{n+1}{2}\right)
\left( j - \frac{n+1}{2}\right) \frac{|i-j|}{2},
\end{equation}
where the term $(i+j)\sigma^2/2$ in Eq.~(\ref{e.<yiyj>}) vanishes considering the
  summation over $i$ and $j$, because $\sum_{i=1}^n[i-(n+1)/2] = 0$. By
Eqs.~(\ref{e.<yiyj>})--(\ref{e.<a^2>}), it follows that the DFA in
Eq.~(\ref{e.F(n)^2}) is rewritten as

\begin{eqnarray}
  \label{e.dfa_general}
  F^2(n)
  &=& \frac{1}{n} \sum_{i=1}^n \sigma^2 i
      - \frac{2}{n} \left[ \sum_{i=1}^n \left( i - \frac{n+1}{2}\right) \langle a \tilde{y}_i \rangle
      + \frac{1}{n} \sum_{i,j=1}^n \langle \tilde{y}_i \tilde{y}_j \rangle  \right]\\
  && +\frac{1}{n} \sum_{i=1}^n \left[ \left( i - \frac{n+1}{2} \right)^2 \langle a^2  \rangle
     +\frac{2}{n} \sum_{j=1}^n \left( i - \frac{n+1}{2} \right) \langle a \tilde{y}_j  \rangle
     +\frac{1 }{n^2} \sum_{j,k=1}^n \langle \tilde{y}_{j} \tilde{y}_{k} \rangle  \right] \\
  &\simeq& \frac{\sigma^2 n}{2}
      +\frac{\lambda}{n^2S_x^2} \sum_{j,k=1}^n \left( j - \frac{n+1}{2}\right)
      \left( k - \frac{n+1}{2}\right) \frac{|k-j|}{2}
      -\frac{\sigma^2n}{2} + \frac{n}{6}\lambda \\
  &\simeq& \frac{\lambda n}{15}, 
\label{dfa-th}
\end{eqnarray}
\end{widetext}
where we used an approximation, $\sum_{j=1}^n j^k \simeq n^{k+1}/(k+1)$ for
$n\to\infty$. Equation (\ref{e.dfa_general}) is valid for any
  point process that has a finite mean and variance. Thus,
  Eq.~(\ref{e.dfa_general}) holds for the Poisson processes, and thus we obtain
  Eq.~(\ref{e.dfa_poisson_coef}).

\section {Asymptotic behavior of the DFA for aftershock sequences}\label{s.theory-dfa-aftershocks}
Here, we evaluate the asymptotic behavior of the DFA for aftershock sequences in the large-$n$ limit. 
The mean number of aftershocks is given by $\langle N(t) \rangle = \int_0^t \lambda_a(t')dt'$. 
Because the derivative of $\langle N(t) \rangle$ tends to zero for $t\to\infty$, deviations from a linear fitting become zero for the large-$t$ limit. 
In other words, the deviation in the first time window in $F(n)$ is significant in the large-$n$ limit. Therefore, 
in the large-$n$ limit, $F(n)$ can be approximately obtained from the first time window:
\begin{equation}
F(n)^2 \cong \chi_0(n)^2 \equiv \frac{1}{mn} \sum_{i=1}^n (y_i - \tilde{y}_i^0)^2.
\label{chi0}
\end{equation}
In the following, we replace $y_i$ with $\langle N(t) \rangle$ to calculate $\chi_0(n)^2$. Using the least mean square method, we have 
a linear function, i.e., $\tilde{y}_i^0 \equiv a i +b$. 

For $p =1$, $\langle N(t) \rangle$ is given by 
\begin{equation}
\langle N(t) \rangle  = K \log \left( \frac{t}{c} + 1 \right).
\end{equation}
Using $\partial \chi_0(n)^2 /\partial a = \partial \chi_0(n)^2 /\partial b =0$, we obtain 
\begin{equation}
a= \frac{3K}{n},~b= K \left[ \log \left(\frac{n}{c}+1 \right) - \frac{5}{2} \right].
\end{equation}
We approximate the sum in Eq.(\ref{chi0}) by the integral:  
\begin{equation}
 \chi_0(n)^2 \cong \frac{1}{T} \int_0^n \left[  K \log \left(\frac{t+c}{n+c}\right) -\frac{3K}{n}t
-\frac{5K}{2} \right]^2 .
\end{equation}
For $n\gg c$, we have
\begin{equation}
F(n) \cong \frac{K}{2} \sqrt{\frac{n}{T}} .
\end{equation}

For $p\ne 1$, $\langle N(t) \rangle$ is given by 
\begin{equation}
\langle N(t) \rangle  = \frac{c^{1-p}K}{1-p} \left[ \left( \frac{t}{c} + 1 \right)^{1-p} - 1 \right].
\end{equation}
In the long-$t$ limit, $\langle N(t) \rangle$  becomes 
\begin{equation}
\langle N(t) \rangle  \sim \frac{K}{1-p} t^{1-p} 
\end{equation}
and 
\begin{equation}
\langle N(t) \rangle  - \frac{c^{1-p}K}{p-1}  \sim - \frac{K}{p-1} t^{1-p} 
\end{equation}
for $p<1$ and $p>1$, respectively. 
In the same calculation as  the above, we have 
\begin{equation}
F(n) \cong  \frac{Kp}{\sqrt{T(3-2p)} (3-p)(2-p)} n^{\frac{3}{2}-p}
\end{equation}
for $p<3/2$ ($p\ne 1$). 


\begin{thebibliography}{46}%
\makeatletter
\providecommand \@ifxundefined [1]{%
 \@ifx{#1\undefined}
}%
\providecommand \@ifnum [1]{%
 \ifnum #1\expandafter \@firstoftwo
 \else \expandafter \@secondoftwo
 \fi
}%
\providecommand \@ifx [1]{%
 \ifx #1\expandafter \@firstoftwo
 \else \expandafter \@secondoftwo
 \fi
}%
\providecommand \natexlab [1]{#1}%
%
\providecommand \bibnamefont  [1]{#1}%
\providecommand \bibfnamefont [1]{#1}%
\providecommand \citenamefont [1]{#1}%
\providecommand \href@noop [0]{\@secondoftwo}%
\providecommand \href [0]{\begingroup \@sanitize@url \@href}%
\providecommand \@href[1]{\@@startlink{#1}\@@href}%
\providecommand \@@href[1]{\endgroup#1\@@endlink}%
\providecommand \@sanitize@url [0]{\catcode `\\12\catcode `\$12\catcode
  `\&12\catcode `\#12\catcode `\^12\catcode `\_12\catcode `\%12\relax}%
\providecommand \@@startlink[1]{}%
\providecommand \@@endlink[0]{}%
\providecommand \url  [0]{\begingroup\@sanitize@url \@url }%
\providecommand \@url [1]{\endgroup\@href {#1}{\urlprefix }}%
\providecommand \urlprefix  [0]{URL }%
%
\providecommand \doibase [0]{http://dx.doi.org/}%
\providecommand \selectlanguage [0]{\@gobble}%
\providecommand \bibinfo  [0]{\@secondoftwo}%
\providecommand \bibfield  [0]{\@secondoftwo}%
%
\providecommand \BibitemOpen [0]{}%
%
%
%
\providecommand \BibitemShut  [1]{\csname bibitem#1\endcsname}%
\let\auto@bib@innerbib\@empty
\bibitem [{\citenamefont {Scher}\ and\ \citenamefont
  {Montroll}(1975)}]{Scher1975}%
  \BibitemOpen
  \bibfield  {author} {\bibinfo {author} {\bibfnamefont {H.}~\bibnamefont
  {Scher}}\ and\ \bibinfo {author} {\bibfnamefont {E.~W.}\ \bibnamefont
  {Montroll}},\ }\href@noop {} {\bibfield  {journal} {\bibinfo  {journal}
  {Phys. Rev. B}\ }\textbf {\bibinfo {volume} {12}},\ \bibinfo {pages} {2455}
  (\bibinfo {year} {1975})}\BibitemShut {NoStop}%
\bibitem [{\citenamefont {Bouchaud}\ and\ \citenamefont
  {Georges}(1990)}]{bouchaud90}%
  \BibitemOpen
  \bibfield  {author} {\bibinfo {author} {\bibfnamefont {J.-P.}\ \bibnamefont
  {Bouchaud}}\ and\ \bibinfo {author} {\bibfnamefont {A.}~\bibnamefont
  {Georges}},\ }\href@noop {} {\bibfield  {journal} {\bibinfo  {journal} {Phys.
  Rep.}\ }\textbf {\bibinfo {volume} {195}},\ \bibinfo {pages} {127} (\bibinfo
  {year} {1990})}\BibitemShut {NoStop}%
\bibitem [{\citenamefont {Bouchaud}(1992)}]{Bouchaud1992}%
  \BibitemOpen
  \bibfield  {author} {\bibinfo {author} {\bibfnamefont {J.-P.}\ \bibnamefont
  {Bouchaud}},\ }\href@noop {} {\bibfield  {journal} {\bibinfo  {journal} {J.
  Phys. I}\ }\textbf {\bibinfo {volume} {2}},\ \bibinfo {pages} {1705}
  (\bibinfo {year} {1992})}\BibitemShut {NoStop}%
\bibitem [{\citenamefont {Monthus}\ and\ \citenamefont
  {Bouchaud}(1996)}]{Monthus1996}%
  \BibitemOpen
  \bibfield  {author} {\bibinfo {author} {\bibfnamefont {C.}~\bibnamefont
  {Monthus}}\ and\ \bibinfo {author} {\bibfnamefont {J.-P.}\ \bibnamefont
  {Bouchaud}},\ }\href@noop {} {\bibfield  {journal} {\bibinfo  {journal} {J.
  Phys. A}\ }\textbf {\bibinfo {volume} {29}},\ \bibinfo {pages} {3847}
  (\bibinfo {year} {1996})}\BibitemShut {NoStop}%
\bibitem [{\citenamefont {Brokmann}\ and\ \citenamefont {{\it et
  al}.}(2003)}]{Brokmann2003}%
  \BibitemOpen
  \bibfield  {author} {\bibinfo {author} {\bibfnamefont {X.}~\bibnamefont
  {Brokmann}}\ and\ \bibinfo {author} {\bibnamefont {{\it et al}.}},\
  }\href@noop {} {\bibfield  {journal} {\bibinfo  {journal} {Phys. Rev. Lett.}\
  }\textbf {\bibinfo {volume} {90}},\ \bibinfo {pages} {120601} (\bibinfo
  {year} {2003})}\BibitemShut {NoStop}%
\bibitem [{\citenamefont {Metzler}\ \emph {et~al.}(2014)\citenamefont
  {Metzler}, \citenamefont {Jeon}, \citenamefont {Cherstvy},\ and\
  \citenamefont {Barkai}}]{Metzler2014}%
  \BibitemOpen
  \bibfield  {author} {\bibinfo {author} {\bibfnamefont {R.}~\bibnamefont
  {Metzler}}, \bibinfo {author} {\bibfnamefont {J.-H.}\ \bibnamefont {Jeon}},
  \bibinfo {author} {\bibfnamefont {A.~G.}\ \bibnamefont {Cherstvy}}, \ and\
  \bibinfo {author} {\bibfnamefont {E.}~\bibnamefont {Barkai}},\ }\href@noop {}
  {\bibfield  {journal} {\bibinfo  {journal} {Phys. Chem. Chem. Phys.}\
  }\textbf {\bibinfo {volume} {16}},\ \bibinfo {pages} {24128} (\bibinfo {year}
  {2014})}\BibitemShut {NoStop}%
\bibitem [{\citenamefont {Omori}(1894)}]{Omori}%
  \BibitemOpen
  \bibfield  {author} {\bibinfo {author} {\bibfnamefont {F.}~\bibnamefont
  {Omori}},\ }\href@noop {} {\bibfield  {journal} {\bibinfo  {journal} {J.
  College Sci. Imp. Univ. Tokyo}\ }\textbf {\bibinfo {volume} {7}},\ \bibinfo
  {pages} {111} (\bibinfo {year} {1894})}\BibitemShut {NoStop}%
\bibitem [{\citenamefont {Utsu}(1970)}]{Utsu1970}%
  \BibitemOpen
  \bibfield  {author} {\bibinfo {author} {\bibfnamefont {T.}~\bibnamefont
  {Utsu}},\ }\href@noop {} {\bibfield  {journal} {\bibinfo  {journal} {J.
  Facul. Sci. Hokkaido Univ. Ser. VII}\ }\textbf {\bibinfo {volume} {3}},\
  \bibinfo {pages} {379 } (\bibinfo {year} {1970})}\BibitemShut {NoStop}%
\bibitem [{\citenamefont {Ogata}(1988)}]{Ogata1988}%
  \BibitemOpen
  \bibfield  {author} {\bibinfo {author} {\bibfnamefont {Y.}~\bibnamefont
  {Ogata}},\ }\href@noop {} {\bibfield  {journal} {\bibinfo  {journal} {J. Am.
  Stat. Assoc.}\ }\textbf {\bibinfo {volume} {83}},\ \bibinfo {pages} {9}
  (\bibinfo {year} {1988})}\BibitemShut {NoStop}%
\bibitem [{\citenamefont {Utsu}(1992)}]{Utsu1992}%
  \BibitemOpen
  \bibfield  {author} {\bibinfo {author} {\bibfnamefont {T.}~\bibnamefont
  {Utsu}},\ }\href@noop {} {\emph {\bibinfo {title} {A review of seismicity (in
  Japanese), in Mathematical Seismology}}},\ edited by\ \bibinfo {editor}
  {\bibfnamefont {M.}~\bibnamefont {Saito}},\ Vol.~\bibinfo {volume} {2}\
  (\bibinfo  {publisher} {Inst. of Stat. Math., Tokyo},\ \bibinfo {year}
  {1992})\BibitemShut {NoStop}%
\bibitem [{\citenamefont {Weigel}\ \emph {et~al.}(2011)\citenamefont {Weigel},
  \citenamefont {Simon}, \citenamefont {Tamkun},\ and\ \citenamefont
  {Krapf}}]{Weigel2011}%
  \BibitemOpen
  \bibfield  {author} {\bibinfo {author} {\bibfnamefont {A.}~\bibnamefont
  {Weigel}}, \bibinfo {author} {\bibfnamefont {B.}~\bibnamefont {Simon}},
  \bibinfo {author} {\bibfnamefont {M.}~\bibnamefont {Tamkun}}, \ and\ \bibinfo
  {author} {\bibfnamefont {D.}~\bibnamefont {Krapf}},\ }\href@noop {}
  {\bibfield  {journal} {\bibinfo  {journal} {Proc. Natl. Acad. Sci. USA}\
  }\textbf {\bibinfo {volume} {108}},\ \bibinfo {pages} {6438} (\bibinfo {year}
  {2011})}\BibitemShut {NoStop}%
\bibitem [{\citenamefont {Yamamoto}\ \emph
  {et~al.}(2014{\natexlab{a}})\citenamefont {Yamamoto}, \citenamefont
  {Akimoto}, \citenamefont {Yasui},\ and\ \citenamefont
  {Yasuoka}}]{Yamamoto2014}%
  \BibitemOpen
  \bibfield  {author} {\bibinfo {author} {\bibfnamefont {E.}~\bibnamefont
  {Yamamoto}}, \bibinfo {author} {\bibfnamefont {T.}~\bibnamefont {Akimoto}},
  \bibinfo {author} {\bibfnamefont {M.}~\bibnamefont {Yasui}}, \ and\ \bibinfo
  {author} {\bibfnamefont {K.}~\bibnamefont {Yasuoka}},\ }\href@noop {}
  {\bibfield  {journal} {\bibinfo  {journal} {Sci. Rep.}\ }\textbf {\bibinfo
  {volume} {4}},\ \bibinfo {pages} {4720} (\bibinfo {year}
  {2014}{\natexlab{a}})}\BibitemShut {NoStop}%
\bibitem [{\citenamefont {Manzo}\ \emph {et~al.}(2015)\citenamefont {Manzo},
  \citenamefont {Torreno-Pina}, \citenamefont {Massignan}, \citenamefont
  {Lapeyre~Jr}, \citenamefont {Lewenstein},\ and\ \citenamefont
  {Parajo}}]{Manzo2015}%
  \BibitemOpen
  \bibfield  {author} {\bibinfo {author} {\bibfnamefont {C.}~\bibnamefont
  {Manzo}}, \bibinfo {author} {\bibfnamefont {J.~A.}\ \bibnamefont
  {Torreno-Pina}}, \bibinfo {author} {\bibfnamefont {P.}~\bibnamefont
  {Massignan}}, \bibinfo {author} {\bibfnamefont {G.~J.}\ \bibnamefont
  {Lapeyre~Jr}}, \bibinfo {author} {\bibfnamefont {M.}~\bibnamefont
  {Lewenstein}}, \ and\ \bibinfo {author} {\bibfnamefont {M.~F.~G.}\
  \bibnamefont {Parajo}},\ }\href@noop {} {\bibfield  {journal} {\bibinfo
  {journal} {Phys. Rev. X}\ }\textbf {\bibinfo {volume} {5}},\ \bibinfo {pages}
  {011021} (\bibinfo {year} {2015})}\BibitemShut {NoStop}%
\bibitem [{\citenamefont {Akimoto}\ \emph {et~al.}(2020)\citenamefont
  {Akimoto}, \citenamefont {Barkai},\ and\ \citenamefont
  {Radons}}]{Akimoto2020}%
  \BibitemOpen
  \bibfield  {author} {\bibinfo {author} {\bibfnamefont {T.}~\bibnamefont
  {Akimoto}}, \bibinfo {author} {\bibfnamefont {E.}~\bibnamefont {Barkai}}, \
  and\ \bibinfo {author} {\bibfnamefont {G.}~\bibnamefont {Radons}},\ }\href
  {\doibase 10.1103/PhysRevE.101.052112} {\bibfield  {journal} {\bibinfo
  {journal} {Phys. Rev. E}\ }\textbf {\bibinfo {volume} {101}},\ \bibinfo
  {pages} {052112} (\bibinfo {year} {2020})}\BibitemShut {NoStop}%
\bibitem [{\citenamefont {Gutenberg}\ and\ \citenamefont {Richter}(1944)}]{GR}%
  \BibitemOpen
  \bibfield  {author} {\bibinfo {author} {\bibfnamefont {B.}~\bibnamefont
  {Gutenberg}}\ and\ \bibinfo {author} {\bibfnamefont {C.~F.}\ \bibnamefont
  {Richter}},\ }\href@noop {} {\bibfield  {journal} {\bibinfo  {journal} {Bull.
  Seismol. Soc. Am.}\ }\textbf {\bibinfo {volume} {34}},\ \bibinfo {pages}
  {185} (\bibinfo {year} {1944})}\BibitemShut {NoStop}%
\bibitem [{\citenamefont {Nanjo}\ \emph {et~al.}(2012)\citenamefont {Nanjo},
  \citenamefont {Hirata}, \citenamefont {Obara},\ and\ \citenamefont
  {Kasahara}}]{nanjo2012decade}%
  \BibitemOpen
  \bibfield  {author} {\bibinfo {author} {\bibfnamefont {K.}~\bibnamefont
  {Nanjo}}, \bibinfo {author} {\bibfnamefont {N.}~\bibnamefont {Hirata}},
  \bibinfo {author} {\bibfnamefont {K.}~\bibnamefont {Obara}}, \ and\ \bibinfo
  {author} {\bibfnamefont {K.}~\bibnamefont {Kasahara}},\ }\href@noop {}
  {\bibfield  {journal} {\bibinfo  {journal} {Geophys. Res. Lett.}\ }\textbf
  {\bibinfo {volume} {39}} (\bibinfo {year} {2012})}\BibitemShut {NoStop}%
\bibitem [{\citenamefont {He}\ \emph {et~al.}(2008)\citenamefont {He},
  \citenamefont {Burov}, \citenamefont {Metzler},\ and\ \citenamefont
  {Barkai}}]{He2008}%
  \BibitemOpen
  \bibfield  {author} {\bibinfo {author} {\bibfnamefont {Y.}~\bibnamefont
  {He}}, \bibinfo {author} {\bibfnamefont {S.}~\bibnamefont {Burov}}, \bibinfo
  {author} {\bibfnamefont {R.}~\bibnamefont {Metzler}}, \ and\ \bibinfo
  {author} {\bibfnamefont {E.}~\bibnamefont {Barkai}},\ }\href@noop {}
  {\bibfield  {journal} {\bibinfo  {journal} {Phys. Rev. Lett.}\ }\textbf
  {\bibinfo {volume} {101}},\ \bibinfo {pages} {058101} (\bibinfo {year}
  {2008})}\BibitemShut {NoStop}%
\bibitem [{\citenamefont {Miyaguchi}\ and\ \citenamefont
  {Akimoto}(2011)}]{Miyaguchi2011}%
  \BibitemOpen
  \bibfield  {author} {\bibinfo {author} {\bibfnamefont {T.}~\bibnamefont
  {Miyaguchi}}\ and\ \bibinfo {author} {\bibfnamefont {T.}~\bibnamefont
  {Akimoto}},\ }\href@noop {} {\bibfield  {journal} {\bibinfo  {journal} {Phys.
  Rev. E}\ }\textbf {\bibinfo {volume} {83}},\ \bibinfo {pages} {031926}
  (\bibinfo {year} {2011})}\BibitemShut {NoStop}%
\bibitem [{\citenamefont {Miyaguchi}\ and\ \citenamefont
  {Akimoto}(2015)}]{Miyaguchi2015}%
  \BibitemOpen
  \bibfield  {author} {\bibinfo {author} {\bibfnamefont {T.}~\bibnamefont
  {Miyaguchi}}\ and\ \bibinfo {author} {\bibfnamefont {T.}~\bibnamefont
  {Akimoto}},\ }\href@noop {} {\bibfield  {journal} {\bibinfo  {journal} {Phys.
  Rev. E}\ }\textbf {\bibinfo {volume} {91}},\ \bibinfo {pages} {010102}
  (\bibinfo {year} {2015})}\BibitemShut {NoStop}%
\bibitem [{\citenamefont {Akimoto}\ and\ \citenamefont
  {Yamamoto}(2016)}]{AkimotoYamamoto2016a}%
  \BibitemOpen
  \bibfield  {author} {\bibinfo {author} {\bibfnamefont {T.}~\bibnamefont
  {Akimoto}}\ and\ \bibinfo {author} {\bibfnamefont {E.}~\bibnamefont
  {Yamamoto}},\ }\href@noop {} {\bibfield  {journal} {\bibinfo  {journal} {J.
  Stat. Mech.}\ }\textbf {\bibinfo {volume} {2016}},\ \bibinfo {pages} {123201}
  (\bibinfo {year} {2016})}\BibitemShut {NoStop}%
\bibitem [{\citenamefont {Wong}\ \emph {et~al.}(2004)\citenamefont {Wong},
  \citenamefont {Gardel}, \citenamefont {Reichman}, \citenamefont {Weeks},
  \citenamefont {Valentine}, \citenamefont {Bausch},\ and\ \citenamefont
  {Weitz}}]{Wong2004}%
  \BibitemOpen
  \bibfield  {author} {\bibinfo {author} {\bibfnamefont {I.~Y.}\ \bibnamefont
  {Wong}}, \bibinfo {author} {\bibfnamefont {M.~L.}\ \bibnamefont {Gardel}},
  \bibinfo {author} {\bibfnamefont {D.~R.}\ \bibnamefont {Reichman}}, \bibinfo
  {author} {\bibfnamefont {E.~R.}\ \bibnamefont {Weeks}}, \bibinfo {author}
  {\bibfnamefont {M.~T.}\ \bibnamefont {Valentine}}, \bibinfo {author}
  {\bibfnamefont {A.~R.}\ \bibnamefont {Bausch}}, \ and\ \bibinfo {author}
  {\bibfnamefont {D.~A.}\ \bibnamefont {Weitz}},\ }\href {\doibase
  10.1103/PhysRevLett.92.178101} {\bibfield  {journal} {\bibinfo  {journal}
  {Phys. Rev. Lett.}\ }\textbf {\bibinfo {volume} {92}},\ \bibinfo {pages}
  {178101} (\bibinfo {year} {2004})}\BibitemShut {NoStop}%
\bibitem [{\citenamefont {Kuno}\ \emph {et~al.}(2000)\citenamefont {Kuno},
  \citenamefont {Fromm}, \citenamefont {Hamann}, \citenamefont {Gallagher},\
  and\ \citenamefont {Nesbitt}}]{kuno2000nonexponential}%
  \BibitemOpen
  \bibfield  {author} {\bibinfo {author} {\bibfnamefont {M.}~\bibnamefont
  {Kuno}}, \bibinfo {author} {\bibfnamefont {D.~P.}\ \bibnamefont {Fromm}},
  \bibinfo {author} {\bibfnamefont {H.~F.}\ \bibnamefont {Hamann}}, \bibinfo
  {author} {\bibfnamefont {A.}~\bibnamefont {Gallagher}}, \ and\ \bibinfo
  {author} {\bibfnamefont {D.~J.}\ \bibnamefont {Nesbitt}},\ }\href@noop {}
  {\bibfield  {journal} {\bibinfo  {journal} {J. Chem. Phys.}\ }\textbf
  {\bibinfo {volume} {112}},\ \bibinfo {pages} {3117} (\bibinfo {year}
  {2000})}\BibitemShut {NoStop}%
\bibitem [{\citenamefont {Corral}(2004)}]{Corral2004}%
  \BibitemOpen
  \bibfield  {author} {\bibinfo {author} {\bibfnamefont {A.}~\bibnamefont
  {Corral}},\ }\href {\doibase 10.1103/PhysRevLett.92.108501} {\bibfield
  {journal} {\bibinfo  {journal} {Phys. Rev. Lett.}\ }\textbf {\bibinfo
  {volume} {92}},\ \bibinfo {pages} {108501} (\bibinfo {year}
  {2004})}\BibitemShut {NoStop}%
\bibitem [{\citenamefont {Abe}\ and\ \citenamefont
  {Suzuki}(2005)}]{abe2005scale}%
  \BibitemOpen
  \bibfield  {author} {\bibinfo {author} {\bibfnamefont {S.}~\bibnamefont
  {Abe}}\ and\ \bibinfo {author} {\bibfnamefont {N.}~\bibnamefont {Suzuki}},\
  }\href@noop {} {\bibfield  {journal} {\bibinfo  {journal} {Physica A}\
  }\textbf {\bibinfo {volume} {350}},\ \bibinfo {pages} {588} (\bibinfo {year}
  {2005})}\BibitemShut {NoStop}%
\bibitem [{\citenamefont {Saichev}\ and\ \citenamefont
  {Sornette}(2006)}]{Saichev2006}%
  \BibitemOpen
  \bibfield  {author} {\bibinfo {author} {\bibfnamefont {A.}~\bibnamefont
  {Saichev}}\ and\ \bibinfo {author} {\bibfnamefont {D.}~\bibnamefont
  {Sornette}},\ }\href {\doibase 10.1103/PhysRevLett.97.078501} {\bibfield
  {journal} {\bibinfo  {journal} {Phys. Rev. Lett.}\ }\textbf {\bibinfo
  {volume} {97}},\ \bibinfo {pages} {078501} (\bibinfo {year}
  {2006})}\BibitemShut {NoStop}%
\bibitem [{\citenamefont {Hasumi}\ \emph {et~al.}(2009)\citenamefont {Hasumi},
  \citenamefont {Akimoto},\ and\ \citenamefont {Aizawa}}]{hasumi2009weibull}%
  \BibitemOpen
  \bibfield  {author} {\bibinfo {author} {\bibfnamefont {T.}~\bibnamefont
  {Hasumi}}, \bibinfo {author} {\bibfnamefont {T.}~\bibnamefont {Akimoto}}, \
  and\ \bibinfo {author} {\bibfnamefont {Y.}~\bibnamefont {Aizawa}},\
  }\href@noop {} {\bibfield  {journal} {\bibinfo  {journal} {Physica A}\
  }\textbf {\bibinfo {volume} {388}},\ \bibinfo {pages} {491} (\bibinfo {year}
  {2009})}\BibitemShut {NoStop}%
\bibitem [{\citenamefont {Tanaka}\ and\ \citenamefont
  {Aizawa}(2017)}]{tanaka2017detailed}%
  \BibitemOpen
  \bibfield  {author} {\bibinfo {author} {\bibfnamefont {H.}~\bibnamefont
  {Tanaka}}\ and\ \bibinfo {author} {\bibfnamefont {Y.}~\bibnamefont
  {Aizawa}},\ }\href@noop {} {\bibfield  {journal} {\bibinfo  {journal} {J.
  Phys. Soc. Jpn}\ }\textbf {\bibinfo {volume} {86}},\ \bibinfo {pages}
  {024004} (\bibinfo {year} {2017})}\BibitemShut {NoStop}%
\bibitem [{\citenamefont {Peng}\ \emph {et~al.}(1994)\citenamefont {Peng},
  \citenamefont {Buldyrev}, \citenamefont {Havlin}, \citenamefont {Simons},
  \citenamefont {Stanley},\ and\ \citenamefont {Goldberger}}]{Peng1994}%
  \BibitemOpen
  \bibfield  {author} {\bibinfo {author} {\bibfnamefont {C.-K.}\ \bibnamefont
  {Peng}}, \bibinfo {author} {\bibfnamefont {S.~V.}\ \bibnamefont {Buldyrev}},
  \bibinfo {author} {\bibfnamefont {S.}~\bibnamefont {Havlin}}, \bibinfo
  {author} {\bibfnamefont {M.}~\bibnamefont {Simons}}, \bibinfo {author}
  {\bibfnamefont {H.~E.}\ \bibnamefont {Stanley}}, \ and\ \bibinfo {author}
  {\bibfnamefont {A.~L.}\ \bibnamefont {Goldberger}},\ }\href {\doibase
  10.1103/PhysRevE.49.1685} {\bibfield  {journal} {\bibinfo  {journal} {Phys.
  Rev. E}\ }\textbf {\bibinfo {volume} {49}},\ \bibinfo {pages} {1685}
  (\bibinfo {year} {1994})}\BibitemShut {NoStop}%
\bibitem [{\citenamefont {Lennartz}\ \emph {et~al.}(2008)\citenamefont
  {Lennartz}, \citenamefont {Livina}, \citenamefont {Bunde},\ and\
  \citenamefont {Havlin}}]{lennartz2008long}%
  \BibitemOpen
  \bibfield  {author} {\bibinfo {author} {\bibfnamefont {S.}~\bibnamefont
  {Lennartz}}, \bibinfo {author} {\bibfnamefont {V.}~\bibnamefont {Livina}},
  \bibinfo {author} {\bibfnamefont {A.}~\bibnamefont {Bunde}}, \ and\ \bibinfo
  {author} {\bibfnamefont {S.}~\bibnamefont {Havlin}},\ }\href@noop {}
  {\bibfield  {journal} {\bibinfo  {journal} {Europhys. Lett.}\ }\textbf
  {\bibinfo {volume} {81}},\ \bibinfo {pages} {69001} (\bibinfo {year}
  {2008})}\BibitemShut {NoStop}%
\bibitem [{\citenamefont {Taqqu}\ \emph {et~al.}(1995)\citenamefont {Taqqu},
  \citenamefont {Teverovsky},\ and\ \citenamefont {Willinger}}]{taqqu95}%
  \BibitemOpen
  \bibfield  {author} {\bibinfo {author} {\bibfnamefont {M.~S.}\ \bibnamefont
  {Taqqu}}, \bibinfo {author} {\bibfnamefont {V.}~\bibnamefont {Teverovsky}}, \
  and\ \bibinfo {author} {\bibfnamefont {W.}~\bibnamefont {Willinger}},\
  }\href@noop {} {\bibfield  {journal} {\bibinfo  {journal} {Fractals}\
  }\textbf {\bibinfo {volume} {03}},\ \bibinfo {pages} {785} (\bibinfo {year}
  {1995})}\BibitemShut {NoStop}%
\bibitem [{\citenamefont {Magdziarz}\ \emph {et~al.}(2009)\citenamefont
  {Magdziarz}, \citenamefont {Weron}, \citenamefont {Burnecki},\ and\
  \citenamefont {Klafter}}]{magdziarz09}%
  \BibitemOpen
  \bibfield  {author} {\bibinfo {author} {\bibfnamefont {M.}~\bibnamefont
  {Magdziarz}}, \bibinfo {author} {\bibfnamefont {A.}~\bibnamefont {Weron}},
  \bibinfo {author} {\bibfnamefont {K.}~\bibnamefont {Burnecki}}, \ and\
  \bibinfo {author} {\bibfnamefont {J.}~\bibnamefont {Klafter}},\ }\href
  {\doibase 10.1103/PhysRevLett.103.180602} {\bibfield  {journal} {\bibinfo
  {journal} {Phys. Rev. Lett.}\ }\textbf {\bibinfo {volume} {103}},\ \bibinfo
  {pages} {180602} (\bibinfo {year} {2009})}\BibitemShut {NoStop}%
\bibitem [{\citenamefont {Cox}(1962)}]{Cox}%
  \BibitemOpen
  \bibfield  {author} {\bibinfo {author} {\bibfnamefont {D.~R.}\ \bibnamefont
  {Cox}},\ }\href@noop {} {\emph {\bibinfo {title} {Renewal theory}}}\
  (\bibinfo  {publisher} {Methuen},\ \bibinfo {address} {London},\ \bibinfo
  {year} {1962})\BibitemShut {NoStop}%
\bibitem [{\citenamefont {Gardner}\ and\ \citenamefont
  {Knopoff}(1974)}]{gardner1974sequence}%
  \BibitemOpen
  \bibfield  {author} {\bibinfo {author} {\bibfnamefont {J.}~\bibnamefont
  {Gardner}}\ and\ \bibinfo {author} {\bibfnamefont {L.}~\bibnamefont
  {Knopoff}},\ }\href@noop {} {\bibfield  {journal} {\bibinfo  {journal} {Bull.
  Seismol. Soc. Am.}\ }\textbf {\bibinfo {volume} {64}},\ \bibinfo {pages}
  {1363} (\bibinfo {year} {1974})}\BibitemShut {NoStop}%
\bibitem [{\citenamefont {Kagan}\ and\ \citenamefont
  {Jackson}(1991)}]{kagan1991long}%
  \BibitemOpen
  \bibfield  {author} {\bibinfo {author} {\bibfnamefont {Y.~Y.}\ \bibnamefont
  {Kagan}}\ and\ \bibinfo {author} {\bibfnamefont {D.~D.}\ \bibnamefont
  {Jackson}},\ }\href@noop {} {\bibfield  {journal} {\bibinfo  {journal}
  {Geophys. J. Int.}\ }\textbf {\bibinfo {volume} {104}},\ \bibinfo {pages}
  {117} (\bibinfo {year} {1991})}\BibitemShut {NoStop}%
\bibitem [{Note1()}]{Note1}%
  \BibitemOpen
  \bibinfo {note} {Japan Meteorological Agency Earthquake Catalog,
  http://evrrss.eri.u-tokyo.ac.jp/tseis/jma1/index.html}\BibitemShut {NoStop}%
\bibitem [{\citenamefont {Ouillon}\ and\ \citenamefont
  {Sornette}(2005)}]{ouillon2005magnitude}%
  \BibitemOpen
  \bibfield  {author} {\bibinfo {author} {\bibfnamefont {G.}~\bibnamefont
  {Ouillon}}\ and\ \bibinfo {author} {\bibfnamefont {D.}~\bibnamefont
  {Sornette}},\ }\href@noop {} {\bibfield  {journal} {\bibinfo  {journal} {J.
  Geophys. Res.}\ }\textbf {\bibinfo {volume} {110}} (\bibinfo {year}
  {2005})}\BibitemShut {NoStop}%
\bibitem [{\citenamefont {Akimoto}\ and\ \citenamefont
  {Aizawa}(2005)}]{akimoto2005large}%
  \BibitemOpen
  \bibfield  {author} {\bibinfo {author} {\bibfnamefont {T.}~\bibnamefont
  {Akimoto}}\ and\ \bibinfo {author} {\bibfnamefont {Y.}~\bibnamefont
  {Aizawa}},\ }\href@noop {} {\bibfield  {journal} {\bibinfo  {journal} {Prog.
  Theor. Phys.}\ }\textbf {\bibinfo {volume} {114}},\ \bibinfo {pages} {737}
  (\bibinfo {year} {2005})}\BibitemShut {NoStop}%
\bibitem [{\citenamefont {Peng}\ \emph {et~al.}(1993)\citenamefont {Peng},
  \citenamefont {Mietus}, \citenamefont {Hausdorff}, \citenamefont {Havlin},
  \citenamefont {Stanley},\ and\ \citenamefont {Goldberger}}]{Peng1993}%
  \BibitemOpen
  \bibfield  {author} {\bibinfo {author} {\bibfnamefont {C.-K.}\ \bibnamefont
  {Peng}}, \bibinfo {author} {\bibfnamefont {J.}~\bibnamefont {Mietus}},
  \bibinfo {author} {\bibfnamefont {J.~M.}\ \bibnamefont {Hausdorff}}, \bibinfo
  {author} {\bibfnamefont {S.}~\bibnamefont {Havlin}}, \bibinfo {author}
  {\bibfnamefont {H.~E.}\ \bibnamefont {Stanley}}, \ and\ \bibinfo {author}
  {\bibfnamefont {A.~L.}\ \bibnamefont {Goldberger}},\ }\href {\doibase
  10.1103/PhysRevLett.70.1343} {\bibfield  {journal} {\bibinfo  {journal}
  {Phys. Rev. Lett.}\ }\textbf {\bibinfo {volume} {70}},\ \bibinfo {pages}
  {1343} (\bibinfo {year} {1993})}\BibitemShut {NoStop}%
\bibitem [{\citenamefont {Koscielny-Bunde}\ \emph {et~al.}(1998)\citenamefont
  {Koscielny-Bunde}, \citenamefont {Bunde}, \citenamefont {Havlin},
  \citenamefont {Roman}, \citenamefont {Goldreich},\ and\ \citenamefont
  {Schellnhuber}}]{Bunde1998}%
  \BibitemOpen
  \bibfield  {author} {\bibinfo {author} {\bibfnamefont {E.}~\bibnamefont
  {Koscielny-Bunde}}, \bibinfo {author} {\bibfnamefont {A.}~\bibnamefont
  {Bunde}}, \bibinfo {author} {\bibfnamefont {S.}~\bibnamefont {Havlin}},
  \bibinfo {author} {\bibfnamefont {H.~E.}\ \bibnamefont {Roman}}, \bibinfo
  {author} {\bibfnamefont {Y.}~\bibnamefont {Goldreich}}, \ and\ \bibinfo
  {author} {\bibfnamefont {H.-J.}\ \bibnamefont {Schellnhuber}},\ }\href
  {\doibase 10.1103/PhysRevLett.81.729} {\bibfield  {journal} {\bibinfo
  {journal} {Phys. Rev. Lett.}\ }\textbf {\bibinfo {volume} {81}},\ \bibinfo
  {pages} {729} (\bibinfo {year} {1998})}\BibitemShut {NoStop}%
\bibitem [{\citenamefont {Havlin}\ \emph {et~al.}(1999)\citenamefont {Havlin},
  \citenamefont {Buldyrev}, \citenamefont {Bunde}, \citenamefont {Goldberger},
  \citenamefont {Ivanov}, \citenamefont {Peng},\ and\ \citenamefont
  {Stanley}}]{havlin1999scaling}%
  \BibitemOpen
  \bibfield  {author} {\bibinfo {author} {\bibfnamefont {S.}~\bibnamefont
  {Havlin}}, \bibinfo {author} {\bibfnamefont {S.}~\bibnamefont {Buldyrev}},
  \bibinfo {author} {\bibfnamefont {A.}~\bibnamefont {Bunde}}, \bibinfo
  {author} {\bibfnamefont {A.}~\bibnamefont {Goldberger}}, \bibinfo {author}
  {\bibfnamefont {P.~C.}\ \bibnamefont {Ivanov}}, \bibinfo {author}
  {\bibfnamefont {C.-K.}\ \bibnamefont {Peng}}, \ and\ \bibinfo {author}
  {\bibfnamefont {H.~E.}\ \bibnamefont {Stanley}},\ }\href@noop {} {\bibfield
  {journal} {\bibinfo  {journal} {Physica A}\ }\textbf {\bibinfo {volume}
  {273}},\ \bibinfo {pages} {46} (\bibinfo {year} {1999})}\BibitemShut
  {NoStop}%
\bibitem [{\citenamefont {Harada}\ \emph {et~al.}(2009)\citenamefont {Harada},
  \citenamefont {Yokogawa}, \citenamefont {Miyaguchi},\ and\ \citenamefont
  {Kori}}]{harada09}%
  \BibitemOpen
  \bibfield  {author} {\bibinfo {author} {\bibfnamefont {T.}~\bibnamefont
  {Harada}}, \bibinfo {author} {\bibfnamefont {T.}~\bibnamefont {Yokogawa}},
  \bibinfo {author} {\bibfnamefont {T.}~\bibnamefont {Miyaguchi}}, \ and\
  \bibinfo {author} {\bibfnamefont {H.}~\bibnamefont {Kori}},\ }\href {\doibase
  10.1529/biophysj.108.139691} {\bibfield  {journal} {\bibinfo  {journal}
  {Biophys. J.}\ }\textbf {\bibinfo {volume} {96}},\ \bibinfo {pages} {255}
  (\bibinfo {year} {2009})}\BibitemShut {NoStop}%
\bibitem [{\citenamefont {Yamamoto}\ \emph
  {et~al.}(2014{\natexlab{b}})\citenamefont {Yamamoto}, \citenamefont
  {Akimoto}, \citenamefont {Hirano}, \citenamefont {Yasui},\ and\ \citenamefont
  {Yasuoka}}]{Yamamoto2014b}%
  \BibitemOpen
  \bibfield  {author} {\bibinfo {author} {\bibfnamefont {E.}~\bibnamefont
  {Yamamoto}}, \bibinfo {author} {\bibfnamefont {T.}~\bibnamefont {Akimoto}},
  \bibinfo {author} {\bibfnamefont {Y.}~\bibnamefont {Hirano}}, \bibinfo
  {author} {\bibfnamefont {M.}~\bibnamefont {Yasui}}, \ and\ \bibinfo {author}
  {\bibfnamefont {K.}~\bibnamefont {Yasuoka}},\ }\href {\doibase
  10.1103/PhysRevE.89.022718} {\bibfield  {journal} {\bibinfo  {journal} {Phys.
  Rev. E}\ }\textbf {\bibinfo {volume} {89}},\ \bibinfo {pages} {022718}
  (\bibinfo {year} {2014}{\natexlab{b}})}\BibitemShut {NoStop}%
\bibitem [{\citenamefont {Paradisi}\ \emph {et~al.}(2012)\citenamefont
  {Paradisi}, \citenamefont {Cesari}, \citenamefont {Donateo}, \citenamefont
  {Contini},\ and\ \citenamefont {Allegrini}}]{paradisi2012scaling}%
  \BibitemOpen
  \bibfield  {author} {\bibinfo {author} {\bibfnamefont {P.}~\bibnamefont
  {Paradisi}}, \bibinfo {author} {\bibfnamefont {R.}~\bibnamefont {Cesari}},
  \bibinfo {author} {\bibfnamefont {A.}~\bibnamefont {Donateo}}, \bibinfo
  {author} {\bibfnamefont {D.}~\bibnamefont {Contini}}, \ and\ \bibinfo
  {author} {\bibfnamefont {P.}~\bibnamefont {Allegrini}},\ }\href@noop {}
  {\bibfield  {journal} {\bibinfo  {journal} {Nonlinear Proc. Geophys.}\
  }\textbf {\bibinfo {volume} {19}},\ \bibinfo {pages} {113} (\bibinfo {year}
  {2012})}\BibitemShut {NoStop}%
\bibitem [{\citenamefont {Akimoto}\ \emph {et~al.}(2018)\citenamefont
  {Akimoto}, \citenamefont {Cherstvy},\ and\ \citenamefont
  {Metzler}}]{Akimoto2018b}%
  \BibitemOpen
  \bibfield  {author} {\bibinfo {author} {\bibfnamefont {T.}~\bibnamefont
  {Akimoto}}, \bibinfo {author} {\bibfnamefont {A.~G.}\ \bibnamefont
  {Cherstvy}}, \ and\ \bibinfo {author} {\bibfnamefont {R.}~\bibnamefont
  {Metzler}},\ }\href {\doibase 10.1103/PhysRevE.98.022105} {\bibfield
  {journal} {\bibinfo  {journal} {Phys. Rev. E}\ }\textbf {\bibinfo {volume}
  {98}},\ \bibinfo {pages} {022105} (\bibinfo {year} {2018})}\BibitemShut
  {NoStop}%
\bibitem [{\citenamefont {Hou}\ \emph {et~al.}(2018)\citenamefont {Hou},
  \citenamefont {Cherstvy}, \citenamefont {Metzler},\ and\ \citenamefont
  {Akimoto}}]{Hou2018}%
  \BibitemOpen
  \bibfield  {author} {\bibinfo {author} {\bibfnamefont {R.}~\bibnamefont
  {Hou}}, \bibinfo {author} {\bibfnamefont {A.~G.}\ \bibnamefont {Cherstvy}},
  \bibinfo {author} {\bibfnamefont {R.}~\bibnamefont {Metzler}}, \ and\
  \bibinfo {author} {\bibfnamefont {T.}~\bibnamefont {Akimoto}},\ }\href@noop
  {} {\bibfield  {journal} {\bibinfo  {journal} {Phys. Chem. Chem. Phys.}\
  }\textbf {\bibinfo {volume} {20}},\ \bibinfo {pages} {20827} (\bibinfo {year}
  {2018})}\BibitemShut {NoStop}%
\bibitem [{\citenamefont {Godr{\`e}che}\ and\ \citenamefont
  {Luck}(2001)}]{God2001}%
  \BibitemOpen
  \bibfield  {author} {\bibinfo {author} {\bibfnamefont {C.}~\bibnamefont
  {Godr{\`e}che}}\ and\ \bibinfo {author} {\bibfnamefont {J.~M.}\ \bibnamefont
  {Luck}},\ }\href@noop {} {\bibfield  {journal} {\bibinfo  {journal} {J. Stat.
  Phys.}\ }\textbf {\bibinfo {volume} {104}},\ \bibinfo {pages} {489} (\bibinfo
  {year} {2001})}\BibitemShut {NoStop}%
\end{thebibliography}
%

\end {document}